\documentclass[11pt,french,british]{article}
\usepackage[T1]{fontenc}
\usepackage[latin1]{inputenc}
\usepackage{a4wide}
\usepackage{babel}
\usepackage{graphics}

\makeatletter

\providecommand{\LyX}{L\kern-.1667em\lower.25em\hbox{Y}\kern-.125emX\@}

\makeatother
\begin{document}

\title{The hard pomeron in soft data}

\author{J.R. Cudell\thanks{
JR.Cudell@ulg.ac.be
}, E. Martynov\thanks{
martynov@bitp.kiev.ua, on leave from the Bogolyubov Institute
for Theoretical Physics, 03143 Kiev, Ukraine
}, O. Selyugin\thanks{
selugin@qcd.theo.phys.ulg.ac.be, on leave from the Bogoliubov Theoretical
Laboratory, JINR, 141980 Dubna, Moscow Region, Russia
}}
\date{{\normalsize Institut de Physique, Université de Liège, 4000 Liège,
Belgium}}
\maketitle
{\centering {\large A. Lengyel\footnote[0]{sasha@len.uzhgorod.ua.}
}\large \par}

\bigskip{}
{\centering Institute of Electron Physics, Universitetska 21, UA-88000
Uzhgorod, Ukraine.\par}

\begin{abstract}
We show that the data for the total cross section and for the real
part of the elastic amplitude indicate the presence of a hard pomeron
in \( \pi p \) and \( Kp \) elastic scattering at \( t=0 \), compatible
with that observed in deep inelastic scattering. We show that such
a hard pomeron is also compatible with \( pp \) and \( \bar{p}p \)
data, provided one unitarises it at high energy.
\end{abstract}

\section{The hard pomeron: what we know}

The existence of a hard singularity in hadronic amplitudes has been
predicted a long time ago~\cite{BFKL}, within the context
of perturbation theory at small-\( x \). It was then shown that a
leading-log\( (s) \) resummation would lead to a square-root branch-cut
in the complex \( j \) plane starting at \[
\alpha ^{ll}_{h}=1+\frac{12\ln 2}{\pi }\alpha _{S}\]
with $\alpha_S$ a fixed value of the strong coupling constant.

Such a fierce singularity has not been seen in data, but it was shown
later that the leading-log\( (s) \) predictions were unstable with
respect to sub-leading resummation \cite{FL, CC}, and that the singularity
was likely to be softer \cite{BLM}. Unfortunately, this result depends
on the algorithm followed to choose the renormalisation scale. 
Nevertheless, the main message is that perturbative QCD leads to a 
strong singularity.

As most of the data have some soft physics intertwined with short-distance
effects, this {}``pure{}'' BFKL pomeron may be transformed into
another object because of long-distance corrections. In fact, it is
possible that such a singularity is already present in HERA data \cite{DL}.
If one assumes that the singularities of hadronic elastic amplitudes
are well approximated by simple poles only, then one needs to introduce
a new singularity, apparently not present in soft cross sections,
to account for the rise of \( F_{2} \) at small \( x \). This new
singularity was taken to be a simple pole, in which case one obtains
a phenomenological estimate of its intercept \cite{DL}: \[
1.39<\alpha _{h}<1.44.\]

\noindent From quasi-elastic vector meson production, one can obtain an estimate
\cite{DL} of the slope of the new trajectory\[
\alpha '\approx 0.1{\rm \: GeV}^{-2}.\]

One of the troublesome properties of this singularity is that it is
manifest only in off-shell photon cross sections. One may argue that,
as standard factorisation theorems do not apply then, one can have
a singularity that is not present in purely hadronic data. It is in
fact unclear whether this singularity should be present in the photon-proton
total cross section, for which factorisation cannot be proven either.
A recent extrapolation \cite{DL} estimates that the ratios of the
soft pomeron to the hard pomeron coupling is given, for the total
\( \gamma p \) cross section, by\begin{equation}
\label{DIScoup}
\frac{g_{hard}}{g_{soft}}\approx 0.002.
\end{equation}
 It is possible however that the hard pomeron coupling is zero in
this case.

So far, no observation of the hard pomeron has been reported
in soft data, although several authors have shown that the inclusion of 
a hard pomeron in soft data is possible \cite{hardinsoft}. 
We shall argue here that such a singularity is in fact a 
necessary ingredient
to obtain a good fit to all forward soft data \( - \) provided that
one uses a simple pole to describe the soft pomeron.

\section{Previous fits to soft data}

A considerable effort \cite{COMPETE02} has recently been
devoted to the reproduction of soft data through analytical fits based
on \( S \)-matrix theory. The main difference between the forms used
concerns the pomeron term, for which three main classes of dependence
in \( s \) have been considered: \( \ln \frac{s}{s_{d}} \), \( \ln
^{2}\frac{s}{s_{t}}+C \),
and simple poles \( \left( \frac{s}{s_{1}}\right) ^{\alpha } \).
Although these three forms for the pomeron work reasonably well in
the description of total cross sections at high energy (\( \sqrt{s}>10 \)
GeV), the simple-pole description fails if the energy threshold is
lowered to \( \sqrt{s}>5 \) GeV, or if the real part of the amplitude
is included, whereas the logarithmic forms achieve a good fit quality
down to 5 GeV. Note that this is rather strange on theoretical grounds,
as one would expect unitarised forms to work better at high-energy.
We show in Table I the results corresponding to those obtained by
the COMPETE collaboration \cite{COMPETE02, RPP}, but with the
updated dataset used in the present study \cite{data}: we consider
all\footnote{Because of the ambiguities linked to nuclear effects, we excluded
cosmic-ray data.} \( pp \), \( \bar{p}p \), \( K^{\pm }p \) and \( \pi ^{\pm }p \)
data for the total cross section and for the \( \rho  \) parameter,
as well as all \( \gamma p \) and \( \gamma \gamma  \) data for
the total cross section.
\begin{table}

{\centering \begin{tabular}{|c|c|c|c|c|}
\hline
\multicolumn{2}{|c}{}&
\multicolumn{3}{|c|}{
\( \chi ^{2} \)/n.o.p.
}\\
\hline

Process
&

 \( N_{p} \)
&

 Simple pole
&

 Dipole
&

 Tripole
\\
\hline

\( \sigma (pp) \)
&

104
&
0.93
&
 0.89
&
 0.88
\\
\hline
\( \sigma (\bar{p}p) \)
&
 59
&
 1.1
&
 1.0
&
 1.2
\\
\hline
\( \sigma (\pi ^{+}p) \)
&
 50
&
 1.4
&
 0.67
&
 0.71
\\
\hline
\( \sigma (\pi ^{-}p) \)
&
 95
&
 0.94
&
 1.0
&
 0.96
\\
\hline
\( \sigma (K^{+}p) \)
&
 40
&
 1.0
&
 0.72
&
 0.71
\\
\hline
\( \sigma (K^{-}p) \)
&
 63
&
 0.73
&
 0.62
&
 0.62
\\
\hline
\( \sigma (\gamma p) \)
&
 41
&
 0.56
&
 0.65
&
 0.61
\\
\hline
\( \sigma (\gamma \gamma ) \)
&
 36
&
 0.88
&
 1.0
&
 0.80
\\
\hline
\( \rho (pp) \)
&
 64
&
 1.9
&
 1.7
&

 1.8
\\
\hline

\( \rho (\bar{p}p) \)
&

 11
&

 0.55
&

 0.44
&

 0.52
\\
\hline

\( \rho (\pi ^{+}p) \)
&

 8
&

 2.7
&

 1.5
&

 1.5
\\
\hline

\( \rho (\pi ^{-}p) \)
&

 30
&

 2.1
&

 1.2
&

 1.1
\\
\hline

\( \rho (K^{+}p) \)
&

 10
&

 0.87
&

 1.1
&

 1.0
\\
\hline

\( \rho (K^{-}p) \)
&

 8
&

 1.7
&

 1.3
&

 0.99
\\
\hline
\hline

all, \( \chi ^{2}_{tot} \)
&

 619
&

 696
&

 590
&

 595
\\
\hline

all, \( \chi ^{2}/\)d.o.f. 
&

 619
&

 1.15
&

 0.98
&

 0.98
\\
\hline
\end{tabular}\par}

\caption{Partial \protect\( \chi ^{2}\protect \) per number of data points
(\protect\( \chi ^{2}\protect \)/n.o.p.) and total \protect\( \chi ^{2}\protect
\)
per degree of freedom (\protect\( \chi ^{2}\protect \)/d.o.f.) for
the COMPETE parametrisations \cite{COMPETE02, RPP}, fitted to the
latest data \cite{data}, for 5~GeV\protect\( <\sqrt{s}<2\protect \)
TeV. }

\end{table}

\bigskip{}

As one can see from Table 1, the main problem of the simple pole fit
stems from the \( Kp \) and \( \pi p \) data, and particularly from
the \( \rho  \) parameter. Hence we want first to re-consider the
treatment of the real part of the amplitude. We have improved the
fit of \cite{COMPETE02, RPP} by including the following sub-leading
effects:

\begin{enumerate}
\item We started with a parametrisation for the imaginary part of the 
asymptotic elastic
amplitude \( ab\rightarrow~ab \). Regge theory predicts that it is
a function of \( \cos \theta
_{t}=\frac{s-m_{a}^{2}-m_{b}^{2}}{2m_{a}m_{b}}=\frac{(s-u)/2}{2m_{a}m_{b}} \),
with \( \theta _{t} \) the scattering angle for the crossed-channel
process. We re-absorbed the denominator in the definition of the couplings,
and then expressed the cross section using exact flux factors, which
for 3 exchanges can be written as:\begin{equation}
\label{fluxfac}
\sigma _{tot}^{(3)}\equiv \frac{1}{2p m_b}\, \left[ \Im mA^{R}_{+}\left(
\frac{s-u}{2}\right) +\Im mA^{S}_{+}\left( \frac{s-u}{2}\right) \mp \Im
mA_{-}\left( \frac{s-u}{2}\right) \right] ,
\end{equation}
with \( p \) the momentum in the laboratory frame \footnote{%
In the \( \gamma \gamma  \) case, \( 2p m_b \) gets replaced by \( s. \)
} of $b$ and the minus sign for the particle.\\
For all models, we use the same parametrisation of the \( C=-1 \)
part for the process \( ap\rightarrow ap \) (\( a=\bar{p} \), \( p \),
\( \pi ^{\pm } \), \( K^{\pm } \)), \begin{equation}
\label{minus}
\Im mA_{-}(s)=M_{a}\left( \frac{s}{s_{1}}\right) ^{\alpha _{-}}
\end{equation}
with \( s_{1}=1 \) GeV\( ^{2} \). For the \( C=+1 \) part, we use
a common Reggeon contribution, and which we allow to be non-degenerate
with the \( C=-1 \) part:\begin{equation}
\label{plus}
\Im mA^{R}_{+}(s)=P_{a}\left( \frac{s}{s_{1}}\right) ^{\alpha _{+}},
\end{equation}
added to a pomeron term from one of the forms corresponding respectively
to a simple, a double and a triple pole:\begin{equation}
\label{pompol}
\Im mA^S_{+}(s)=S_{a}\left( \frac{s}{s_{1}}\right) ^{\alpha _{o}},
\end{equation}
\begin{equation}
\label{pomd}
\Im mA^S_{+}(s)=D_{a}s\ln \frac{s}{s_{d}},
\end{equation}
\begin{equation}
\label{pomt}
\Im mA^S_{+}(s)=T_{a}s\ln ^{2}\frac{s}{s_{t}}+T'_{a}s
\end{equation}

\item We have fully applied the factorisation constraints in the \( \gamma
\gamma  \)
case: there the couplings \( g \) (standing for \( M \), \( P \)
or \( S \)) of each simple pole can be directly obtained from the
\( pp \) and the \( \gamma p \) fits through the relation \( g_{\gamma \gamma
}=\left( g_{\gamma p}\right) ^{2}/g_{pp} \),
and the couplings of multiple singularities obey more complicated
relations \cite{CMS}.
\item For the derivation of the real part, we used three levels of
sophistication:

\begin{enumerate}
\item Derivative dispersion relations (DDR) \cite{DDR} 
without a subtraction constant. This corresponds to the fit performed in
\cite{COMPETE02, RPP}, but with the exact flux factors and arguments
of Eq. (\ref{fluxfac}).
\item DDR with a free subtraction constant. Because the crossing-even part
of the amplitude rises with energy, one must perform a subtraction,
and the value of the real part at the subtraction point is unknown.
We keep it and fit to it.
\item Integral dispersion relations (IDR) for the analytic parametrisation,
from the threshold \( \sqrt{s_{0}}=m_{a}+m_{b} \). If one takes the
threshold to be zero, the IDR is equivalent to the DDR. However, as
the threshold is nonzero, there is a small correction due to this
shift.
\item IDR for the analytic parametrisation down to \( \sqrt{s}=5 \) GeV,
and to a fit of the data from \( \sqrt{s_{0}} \) to 5 GeV, shown
in Fig. 1. As the analytic forms (\ref{fluxfac})-(\ref{pomt}) do not
reproduce the total cross section data below 5 GeV, we do not use
them there, but instead perform a multi-parameter fit of the total
cross section, shown in Fig. 1. Hence the input below the minimum
energy where the fit is applicable is determined by the data themselves.
It must be emphasised that the details of the low-energy fit have
very little influence on the global fit (see Table 2), mainly because
most of the effects can be re-absorbed in the value of the subtraction
constant.\\
\\
\\
\resizebox*{0.7\textwidth}{!}{\includegraphics{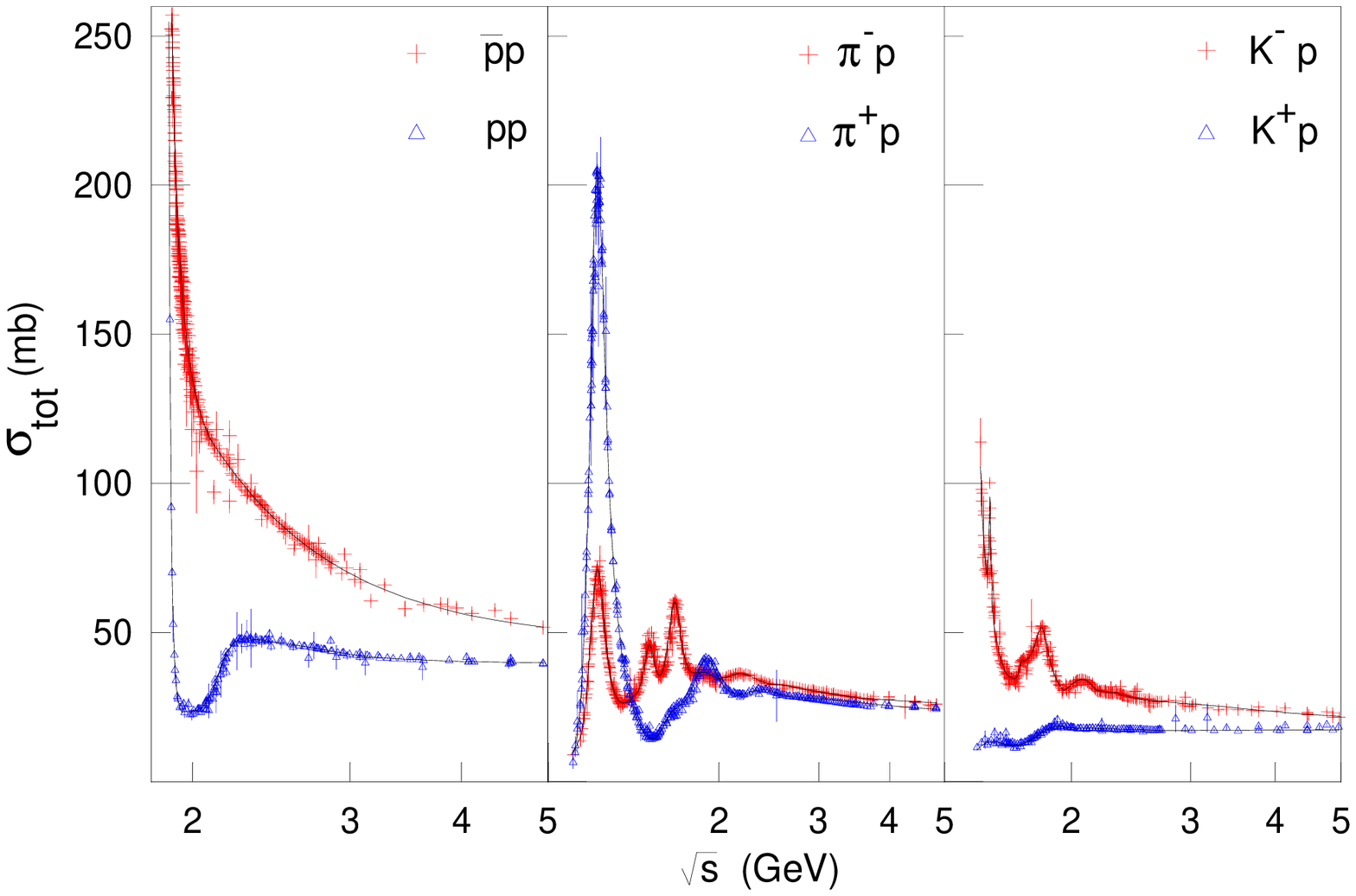}} \\
Figure 1: Fit to low-energy data used in integral dispersion relations.
\end{enumerate}
\end{enumerate}
The formula that we shall be using in this paper (except when otherwise
indicated) for the \( \rho  \) parameter, i.e. the ratio of the real
to the imaginary part of the elastic \( ap \) and \( \bar{a}p \)
amplitudes, {corresponds to case (d) and can
be written\begin{equation}
\label{eq:final dr}
\rho _{\pm }\, \sigma _{\pm }=\frac{R_{ap}}{p}+\frac{E}{\pi p}{\textrm{P}}\int
_{m_{a}}^{\infty }\left[ \frac{\sigma _{\pm }}{E'(E'-E)}-\frac{\sigma _{\mp
}}{E'(E'+E)}\right] p'\, dE'
\end{equation}
}where the \( + \) sign refers to the process \( ap\rightarrow ap \)
and the \( - \) sign to \( \bar{a}p\rightarrow \bar{a}p \), \( E \)
is the energy in the proton rest frame, P indicates that we have to
do a principal-part integral, \( R_{ap} \) is the subtraction constant,
and the fit of Fig. 1 is used for \( \sqrt{s}\leq 5 \) GeV.

The only possible improvement which we have not implemented is the
inclusion of bound-state contributions and the continuation of the fit
to unphysical thresholds. However, at high energy, the main effect
of these corrections can be re-absorbed in the subtraction constant,
leaving a contribution of order \( 1/s \) to the real part. In fact,
the values of the \( \chi ^{2} \)/d.o.f. of the fourth and fifth columns
of Table 1 (cases (b) and (c)) are very similar, precisely because
of this: the shift of the threshold, in this case from \( 0 \) to
\( 2m_{p} \), can be re-absorbed into the subtraction constant. The
resulting values of the \( \chi ^{2} \)/d.o.f. are shown in Table
2, for the simple-pole fit (and for cases (a) to (d)), as well as
for the log and log\( ^{2} \) fits (for case (d)).

\begin{table}

{\centering {\small \begin{tabular}{|c|c|c|c|c|c|c|c|}
\hline
\multicolumn{2}{|c|}{}&
\multicolumn{4}{c|}{
Simple pole
}&

 Dipole
&

 Tripole
\\
\hline

Process
&

 \( N_{p} \)
&

 (a)
&

 (b)
&

 (c)
&

 (d)
&

 (d )
&

 (d)
\\
\hline

\( \sigma (pp) \)
&

 104
&

0.93
&

 1.1
&

 1.1
&

 1.1
&

0.88
&

 0.87
\\
\hline

\( \sigma (\bar{p}p) \)
&

 59
&

 1.0
&

 0.91
&

 0.91
&

 0.88
&

 0.94
&

 0.94
\\
\hline

\( \sigma (\pi ^{+}p) \)
&

 50
&

 1.4
&

 1.2
&

 1.2
&

 1.2
&

 0.68
&

 0.68
\\
\hline

\( \sigma (\pi ^{-}p) \)
&

 95
&

 0.95
&

 0.92
&

 0.92
&

 0.92
&

 0.97
&

 0.97
\\
\hline

\( \sigma (K^{+}p) \)
&

 40
&

 1.0
&

0.96
&

 0.96
&

 0.97
&

 0.73
&

 0.71
\\
\hline

\( \sigma (K^{-}p) \)
&

 63
&

 0.72
&

 0.73
&

 0.73
&

 0.73
&

 0.62
&

 0.61
\\
\hline

\( \sigma (\gamma p) \)
&

 41
&

 0.56
&

 0.56
&

 0.56
&

 0.56
&

 0.58
&

 0.54
\\
\hline

\( \sigma (\gamma \gamma ) \)
&

 36
&

 0.88
&

 0.88
&

 0.88
&

 0.88
&

 0.80
&

 0.73
\\
\hline

\( \rho (pp) \)
&

 64
&

 1.9
&

 1.6
&

 1.6
&

 1.6
&

 1.6
&

 1.7
\\
\hline

\( \rho (\bar{p}p) \)
&

 11
&

 0.49
&

 0.40
&

 0.40
&

 0.40
&

 0.39
&

 0.42
\\
\hline

\( \rho (\pi ^{+}p) \)
&

 8
&

 2.7
&

 2.9
&

 2.9
&

 2.9
&

 1.8
&

 1.8
\\
\hline

\( \rho (\pi ^{-}p) \)
&

 30
&

 2.2
&

 1.9
&

 1.9
&

 1.9
&

 1.0
&

 1.0
\\
\hline

\( \rho (K^{+}p) \)
&

 10
&

 0.91
&

 0.70
&

 0.70
&

 0.70
&

 0.57
&

 0.60
\\
\hline

\( \rho (K^{-}p) \)
&

 8
&

 1.7
&

 1.7
&

 1.7
&

 1.7
&

 1.2
&

 1.0
\\
\hline
\hline

all, \( \chi ^{2}_{tot} \)
&

 619
&

 694
&

661
&

 661
&

 661
&

564
&

 558
\\
\hline

all, \( \chi ^{2}/ \)d.o.f.
&

 619
&

 1.15
&

 1.10
&

 1.10
&

 1.10
&

 0.94
&

 0.93
\\
\hline
\end{tabular}}\small \par}

\caption{{\small Values of the \protect\( \chi ^{2}\protect \)/n.o.p. for
the new parametrisations: (a) the standard (analytic) fit, based on
DDR, with the flux and variables of Eq. (\ref{fluxfac}) and without
subtraction constants; (b) the same fit with subtraction constants;
(c) fit with \protect\( \rho \protect \) calculated by the IDR, using
the high-energy parametrisation from the thresholds; (d) fit of
the high-energy parametrisation with
IDR, using a fixed parametrisation of the cross
section data below \protect\( \sqrt{s}=\protect \)5 GeV.}}

\end{table}

Although the various effects detailed above significantly improve the quality
of the fit, they also improve the dipole and tripole fits, and a simple-pole
pomeron still does not seem acceptable. The only possibility left
to keep this model is to introduce extra singularities and check whether
they can lower the \( \chi ^{2} \)/d.o.f. sufficiently.

\section{The hard pomeron pole}

In fact, we tried to improve the quality of the simple-pole fit by
further lifting the degeneracy of sub-leading vector meson trajectories:
extrapolating hadroscopic data to \( M^{2}=0 \) leads to the conclusion
that the \( f \) intercept is higher than the \( a_{2} \) intercept
\cite{nondeg}. As a first step\footnote{%
as in principle one would have to decouple the \( a_{2} \) from some
of the processes considered here.
}, we simply added one \( C=+1 \) trajectory to the fit, and left
its couplings free (and imposed the corresponding factorisation properties
for the \( \gamma \gamma  \) cross section). This improved the \( \chi ^{2} \)
considerably, and made it comparable to that of the other parametrisations:
Table 3 shows the quality of the fit if one introduces a new \( C=+1 \)
singularity, so that the expression of the cross section now contains
four terms: \begin{equation}
\label{hardpompole}
\sigma _{tot}^{(4)}=\sigma _{tot}^{(3)}+\frac{1}{2p m_b}\Im mA^{H}_{+}
\left({s-u\over 2}\right)
\end{equation}
with\begin{equation}
\label{hardpole}
\Im mA^{H}_{+}(s)=H_{a}\left( \frac{s}{s_{1}}\right) ^{\alpha _{h}}
\end{equation}
with again \( s_{1}=1 \) GeV\( ^{2} \).
\begin{table}

{\centering {\small \begin{tabular}{|c|c|c|c|c|}
\hline
\multicolumn{2}{|c|}{}&

soft pole
&

soft+hard
&

 soft simple pole+
\\
\multicolumn{2}{|c|}{}&

only
&
simple poles&

unitarised hard pole
\\
\hline

Process
&

 \( N_{p} \)
&

 (d)
&

 (d)
&

 (d)
\\
\hline

\( \sigma (pp) \)
&

 104
&

 1.1
&

 0.87
&

 0.87
\\
\hline

\( \sigma (\bar{p}p) \)
&

 59
&

 0.88
&

 0.92
&

 0.92
\\
\hline

\( \sigma (\pi ^{+}p) \)
&

 50
&

 1.2
&

 0.70
&

 0.69
\\
\hline

\( \sigma (\pi ^{-}p) \)
&

 95
&

 0.92
&

 0.93
&

 0.95
\\
\hline

\( \sigma (K^{+}p) \)
&

 40
&

 0.97
&

 0.72
&

 0.72
\\
\hline

\( \sigma (K^{-}p) \)
&

 63
&

 0.73
&

 0.61
&

 0.61
\\
\hline

\( \sigma (\gamma p) \)
&

 41
&

 0.56
&

 0.54
&

 0.56
\\
\hline

\( \sigma (\gamma \gamma ) \)
&

 36
&

 0.88
&

 0.70
&

 0.82
\\
\hline

\( \rho (pp) \)
&

 64
&

 1.6
&

 1.7
&

 1.7
\\
\hline

\( \rho (\bar{p}p) \)
&

 11
&

 0.40
&

 0.41
&

0.40
\\
\hline

\( \rho (\pi ^{+}p) \)
&

 8
&

 2.9
&

 1.6
&

 1.7
\\
\hline

\( \rho (\pi ^{-}p) \)
&

 30
&

 1.9
&

 1.0
&

 1.0
\\
\hline

\( \rho (K^{+}p) \)
&

 10
&

 0.70
&

 0.62
&

 0.60
\\
\hline

\( \rho (K^{-}p) \)
&

 8
&

 1.7
&

 0.98
&

 1.0
\\
\hline
\hline

all, \( \chi ^{2}_{tot} \)
&

 619
&

 661
&

 551
&

 557
\\
\hline

all, \( \chi ^{2}/ \)d.o.f.
&

 619
&

 1.10
&

 0.924
&

 0.933
\\
\hline
\end{tabular}}\small \par}

\caption{{\small The values of \protect\( \chi ^{2}/\protect \)n.o.p., for
\protect\( 5\protect \) GeV\protect\( < \sqrt{s}< 2\protect \)
TeV, as in Table 2 (third column), if we introduce a new pole with
positive charge parity (fourth column, Eq. (\ref{hardpompole})) and
if we unitarise it (fifth column, Eq. (\ref{hardpomunit}))}}

\end{table}

However, this trajectory did not choose an intercept compatible with
that of a Reggeon, but rather settled on an intercept of 1.45, very
close to the one already observed by Donnachie and Landshoff in DIS.
Furthermore, if we fit to Tevatron energies, the trajectory couples
to \( \pi p \) and \( Kp \) processes, but seems absent in \( pp \)
and \( \bar{p}p \).

This is easy to understand if one notices that the \( \pi p \) and
\( Kp \) data have a maximum energy of the order of \( \sqrt{s}=100 \)
GeV. A hard pomeron, if present in soft data, will certainly have
to be unitarised at very large energies \( - \)we shall come back
to this point later\( - \). In fact, the extrapolation of the fit
with 4 poles of Eqs. (\ref{hardpompole}, \ref{hardpole}) gives \( \pi p \)
and \( Kp \) total cross sections much bigger than \( pp \) at the
Tevatron: as it was not unitarised, the fit chose to turn off the
hard pomeron contribution in \( pp \) and \( \bar{p}p \), whereas
the couplings to \( \pi p \) and \( Kp \) were non negligible. This
zero coupling explains in fact why this contribution has been overlooked
before \cite{preCOMPETE}.
\begin{table}

{\centering \begin{tabular}{|c|c|c|c|c|}
\hline
&
\multicolumn{2}{c|}{
 soft+hard poles
}&
\multicolumn{2}{c|}{soft pole+ unitarised hard}\\
\hline

Parameters
&

 value
&

 error
&

 value
&

 error
\\
\hline

\( \alpha _{o} \)
&

 1.0728
&

 0.0008
&

 1.0728
&

 fixed
\\
\hline
\( S_{p} \)&

 56.2
&

 0.3
&

55
&

1
\\
\hline

\( S_{\pi } \)
&

 32.7
&

 0.2
&

 31.5
&

 0.9
\\
\hline
\( S_{K} \)&

 28.3
&

 0.2
&

 27.4
&

 0.8
\\
\hline
\( S_{\gamma } \)&

 0.174
&

 0.002
&

0.174
&

0.003
\\
\hline

\( \alpha _{h}(0) \)
&

 1.45
&

 0.01
&

 1.45
&

 fixed
\\
\hline
\( G_{p} \)&

 --
&

 --
&

 0.18
&

0.06
\\
\hline

\( G_{\gamma } \)
&

 --
&

 --
&

 6\( \times 10^{-9} \)
&

 1.5\( \times 10^{-8} \)
\\
\hline
\( H_{p} \)&

 0.10
&

 0.02
&

0.17
&

 0.05
\\
\hline
\( H_{\pi } \)&

 0.28
&

 0.03
&

0.43
&

 0.08
\\
\hline
\( H_{K} \)&

 0.30
&

 0.03
&

0.42
&

0.07
\\
\hline
\( H_{\gamma } \)&

 0.0006
&

 0.0002
&

 0.0005
&

0.0002
\\
\hline

\( \alpha _{+}(0) \)
&

 0.608
&

 0.003
&

0.62
&

0.02
\\
\hline
\( P_{p} \)&

 158
&

 2
&

 157
&

 5
\\
\hline
\( P_{\pi } \)&

 78
&

 1
&

 80
&

2
\\
\hline

\( P_{K} \)
&

 46
&

 1
&

47
&

 2
\\
\hline
\( P_{\gamma } \)&

 0.28
&

 0.01
&

0.28
&

0.01
\\
\hline

\( \alpha _{-}(0) \)
&

 0.473
&

 0.008
&

0.47
&

0.01
\\
\hline
\( M_{p} \)&

 79
&

 3
&

79
&

3
\\
\hline
\( M_{\pi } \)&

 14.2
&

 0.5
&

14.3
&

0.6
\\
\hline
\( M_{K} \)&

 32
&

 1
&

 32
&

1
\\
\hline
\( R_{pp} \)&

-164
&

 33
&

-163
&

34
\\
\hline
\( R_{p\pi } \)&

-96
&

 21
&

-86
&

21
\\
\hline
\( R_{pK} \)&

 3
&

 26
&

8
&

 26
\\
\hline
\end{tabular}\par}

\caption{Parameters obtained in the fits. The second and third columns give
the parameters and errors of the fit with a hard pole, Eq. (\ref{hardpompole})
for \protect\( \sqrt{s}\protect \) from 5 to 100 GeV, the fourth
and fifth columns give the parameters of a unitarised fit, Eq.
(\ref{hardpomunit})
for 5 GeV\protect\( <\sqrt{s}<2\protect \) TeV. }
\end{table}

Before considering a possible unitarisation scheme, we show in the 
second and third columns of Table
4 the results of a fit for 5 GeV \( <\sqrt{s}< \)100 GeV. The only
difference with the global fit of Table 3 is that the \( \bar{p}p \)
and \( pp \) data do not force the coupling of the hard pomeron to
be zero anymore. Several comments are in order:

\begin{enumerate}
\item The main improvement, as seen from the partial \( \chi ^{2} \) of
Table 3, is in \( \sigma _{\pi ^{+}p} \), \( \sigma _{K^{+}p} \)
and in \( \rho _{\pi ^{+}p} \), \( \rho _{\pi ^{-}p} \) and \( \rho _{K^{-}p}
\).
We show in Figs. 2 and 3 the curves corresponding to these quantities,
where the effect of the hard pomeron can be clearly seen. {}Furthermore,
all processes but two (\( \rho _{pp} \) and \( \rho _{\pi ^{+}p} \))
can now be simultaneously described with a \( \chi ^{2}/\)n.o.p.\(\leq 1 \).\\
\\
\centerline{{ }\resizebox*{0.4\textwidth}{!}{\includegraphics{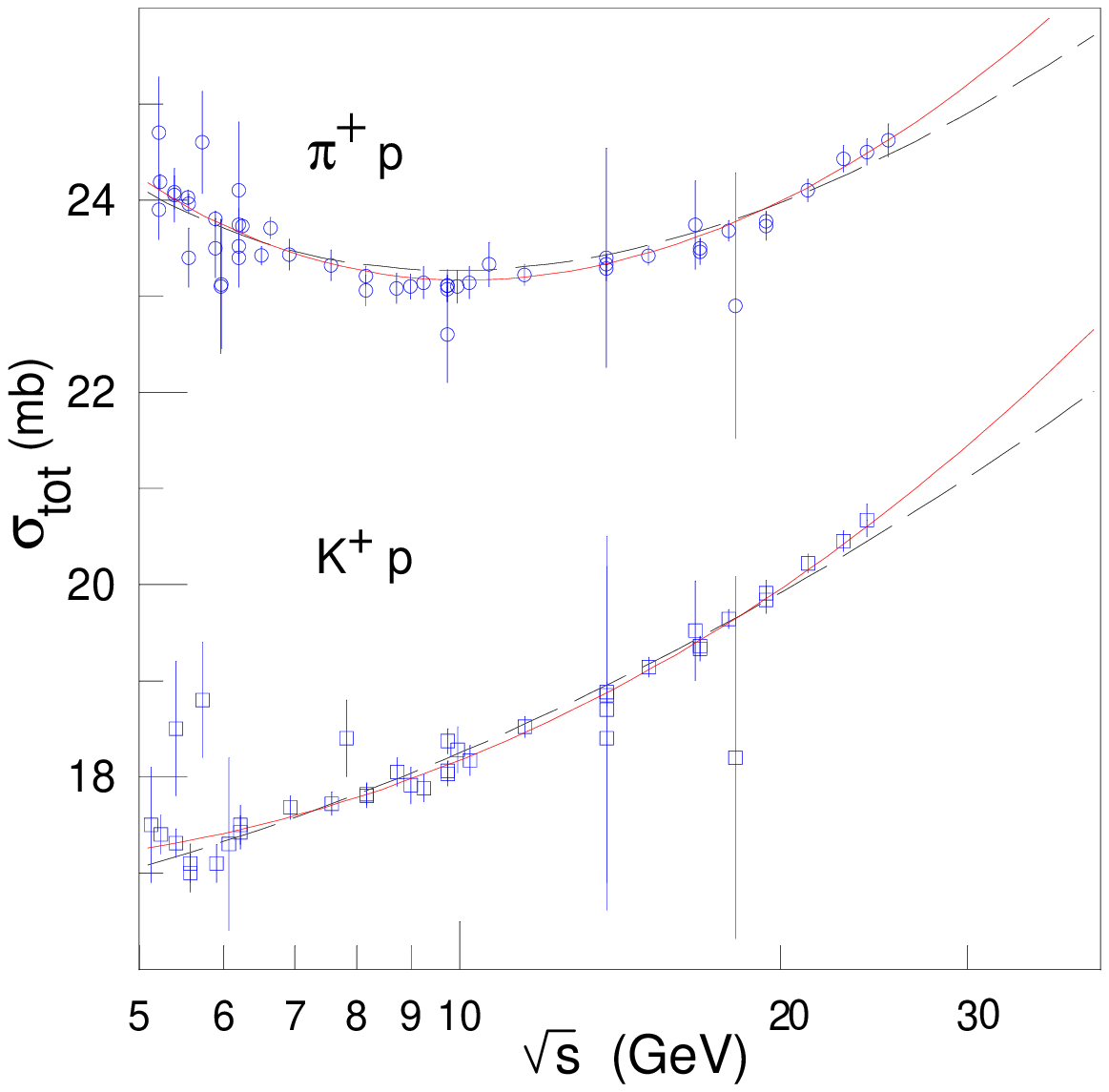}}
{}~~\resizebox*{0.4\textwidth}{!}{\includegraphics{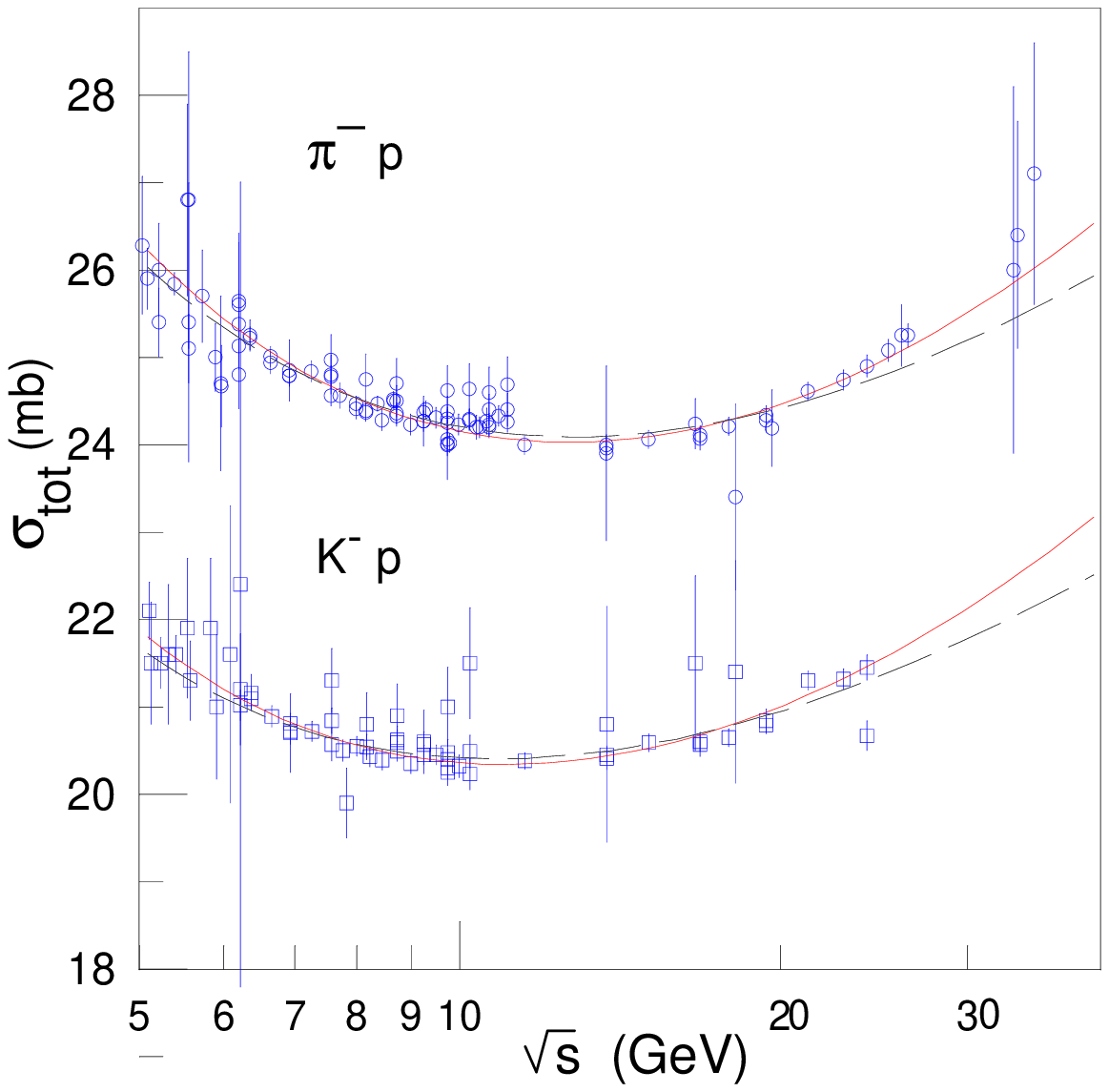}} }\\
Figure 2: Difference between total cross sections fitted with (plain)
and without (dashed) a hard pomeron, assuming all singularities are 
simple poles.\\
\\
\\
\centerline{\resizebox*{0.4\textwidth}{!}{\includegraphics{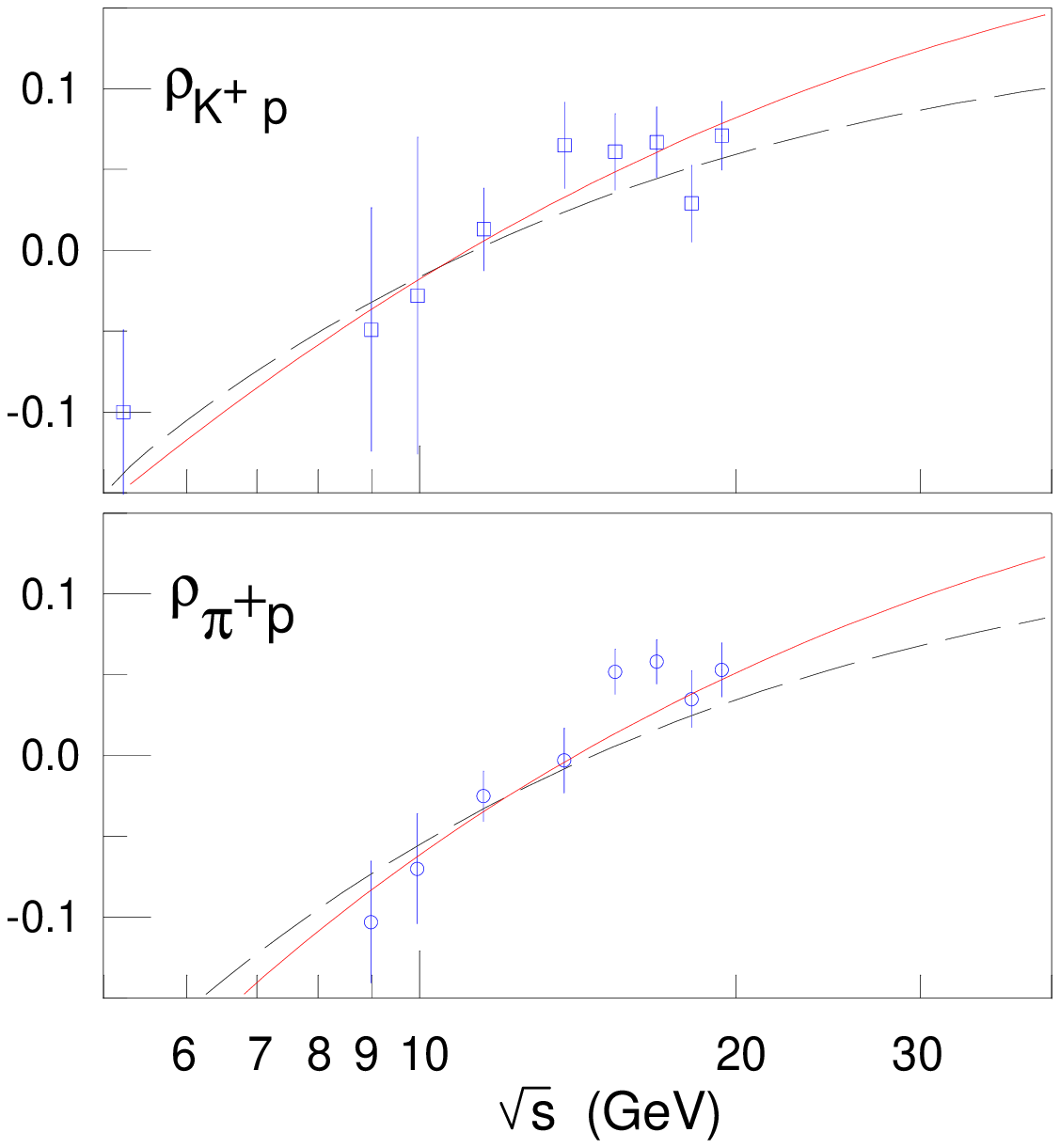}}
\resizebox*{0.4\textwidth}{!}{\includegraphics{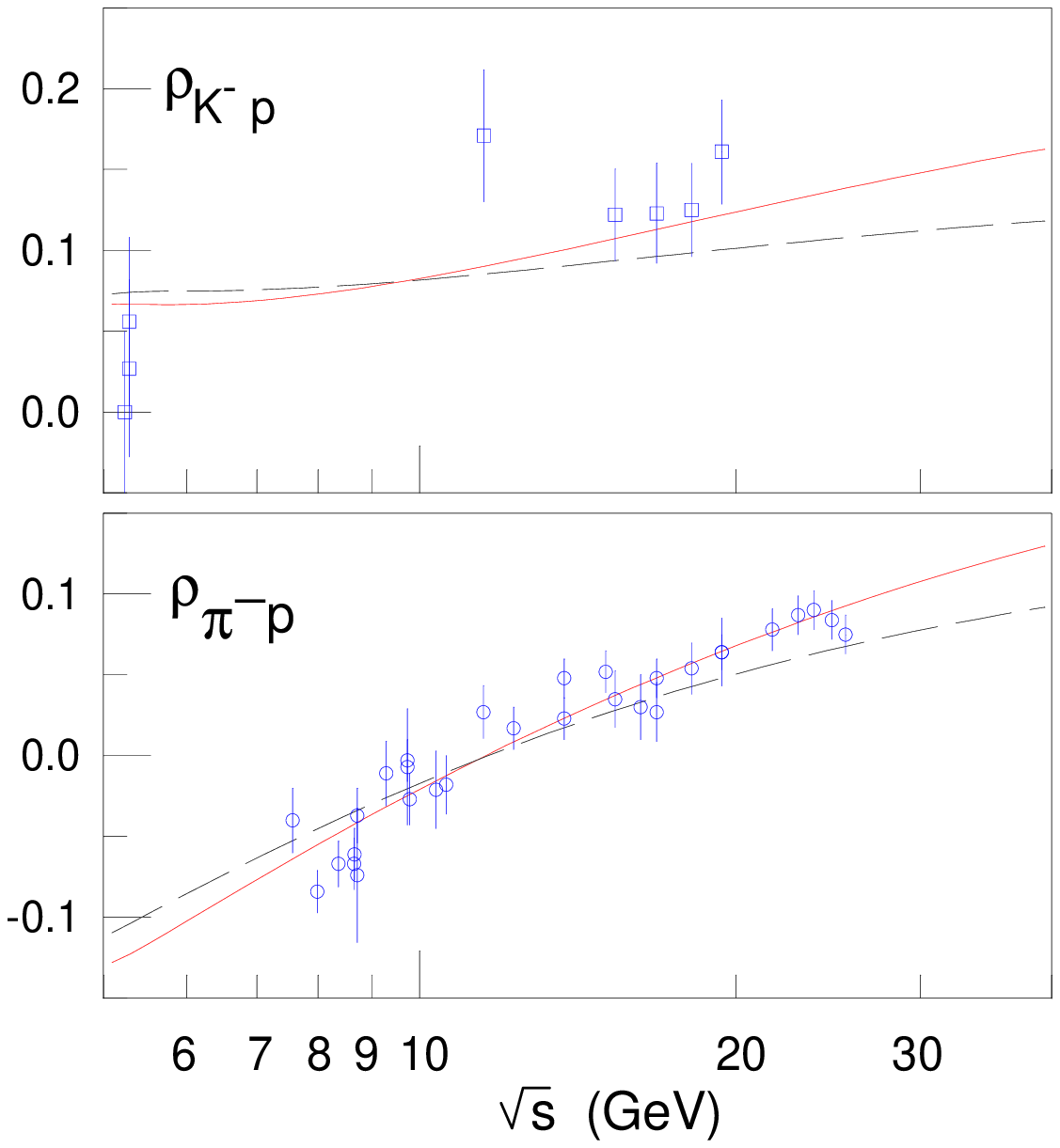}} }
\noindent Figure 3: Difference between \( \rho  \) values fitted with (plain)
and without (dashed) a hard pomeron, assuming all singularities are 
simple poles.\\~\\
\item The value of the hard pomeron intercept is evaluated to be\[
\alpha _{h}=1.45\pm 0.01\]
and is very close to the value obtained in DIS \cite{DL}, and more recently
in $\Upsilon$ photoproduction \cite{Tandler}.
\item The value of the soft pomeron intercept becomes slightly lower than
estimated by Donnachie and Landshoff;
\item The ratio of the coupling of the hard pomeron to the soft one varies
from 0.2\% in \( pp \) and 0.35\% in \( \gamma p \) to 1\% in \( \pi p \)
and \( Kp \). This is compatible with the estimate (\ref{DIScoup})
of \cite{DL}. 
It indicates however that the coupling mechanism of
the hard pomeron must be very different from that of the soft pomeron.
Note however that it is possible to reduce the hard pomeron
coupling to a much smaller value if one does not limit the upper energy of
the fit \cite{DLnew}. 
\item From the values of the coupling and of the intercept, and assuming
a slope \( B=4 \)~GeV\( ^{-2} \) for the proton form factor, and
slopes of \( 0.25 \) GeV\( ^{-2} \) for the soft pomeron and of
\( 0.1 \)~GeV\( ^{-2} \) for the hard pomeron, one can estimate
that the {}``Black-disk{}'' limit will be reached around \( \sqrt{s}=400 \)
GeV. Hence it is likely that if we limit the fit to 100 GeV, we do
see the {}``bare{}'' singularity;
\item Although the hard pomeron has a large intercept, its contribution
to the amplitude remains small because its coupling is tiny. We show
in Fig. 4 the relative contribution of the various terms to the total
cross section. At 100 GeV, the hard pomeron contributes 6\% to the
total cross section. Hence it is possible that it remains hidden,
even in the differential elastic cross section.\\
\\
\\
\centerline{\resizebox*{0.5\textwidth}{!}{\includegraphics{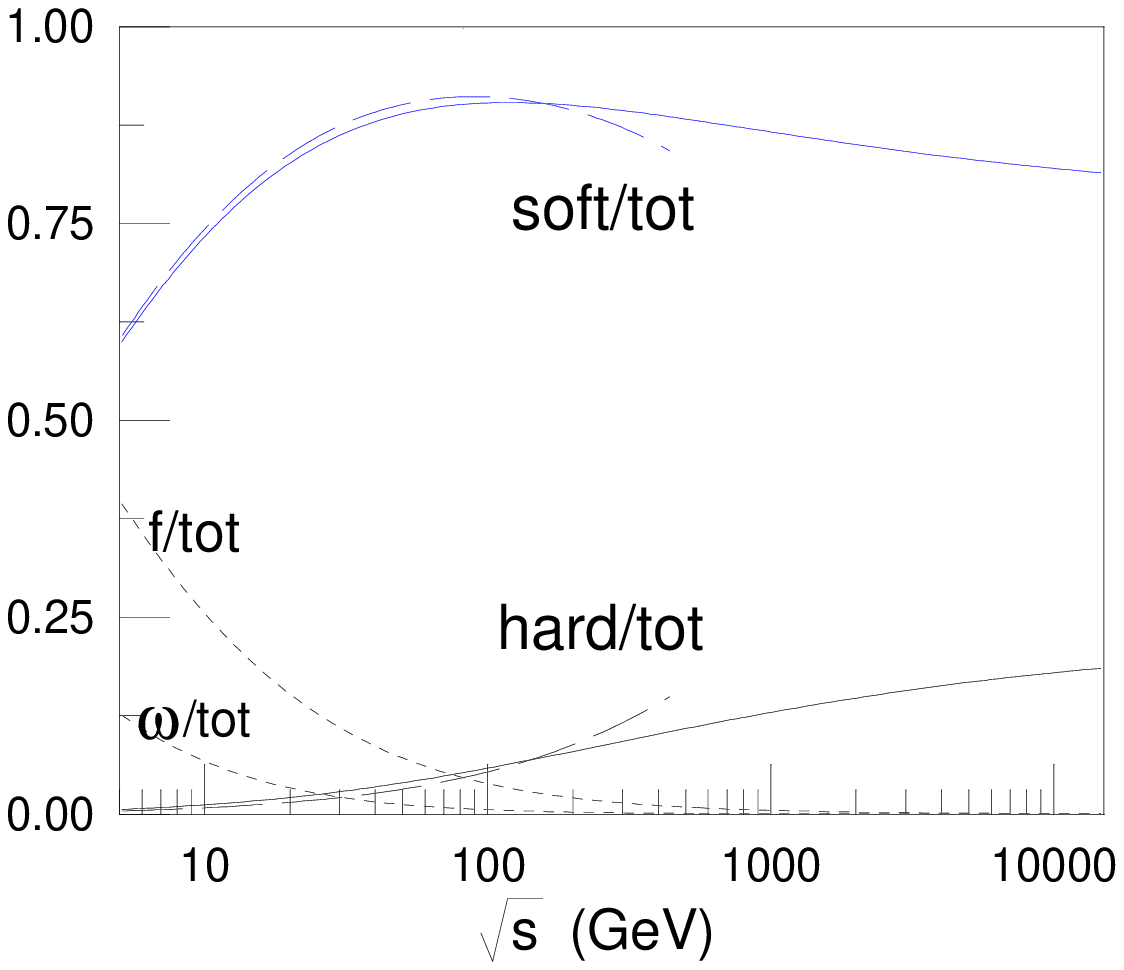}} }\\
Figure 4: Relative contribution of the various terms of the amplitude,
compared with the \( C=+1 \) part of the amplitude ({}``tot{}'')
in the \( pp \) case. The dashed curve is for a hard pole, and the
plain curves for the unitarised form.\\~\\
\item If the hard pomeron is present both in \( pp \) and \( \gamma p \)
scattering, then one can predict the \( \gamma \gamma  \) cross section
through factorisation relations for the couplings of each trajectory.
This leads to the curves shown in Fig. 5, which are thus parameter-free
in the \( \gamma \gamma  \) case. Hence having a hard pomeron does not
necessarily mean that the  \( \gamma \gamma  \) cross section will increase 
faster than in the $\gamma p$ case. Of course, it would also be possible 
to accommodate a faster increase by reducing the hard pomeron coupling
in the pp case (see \cite{DLnew} for such an alternative). Note also
that it is possible to accomodate the $pp$, $\gamma p$ and $\gamma \gamma$
data through factorisation without a hard component \cite{CMS,Block}. 

\end{enumerate}
\vspace{0.3cm}
{\centering \resizebox*{0.45\textwidth}{!}{\includegraphics{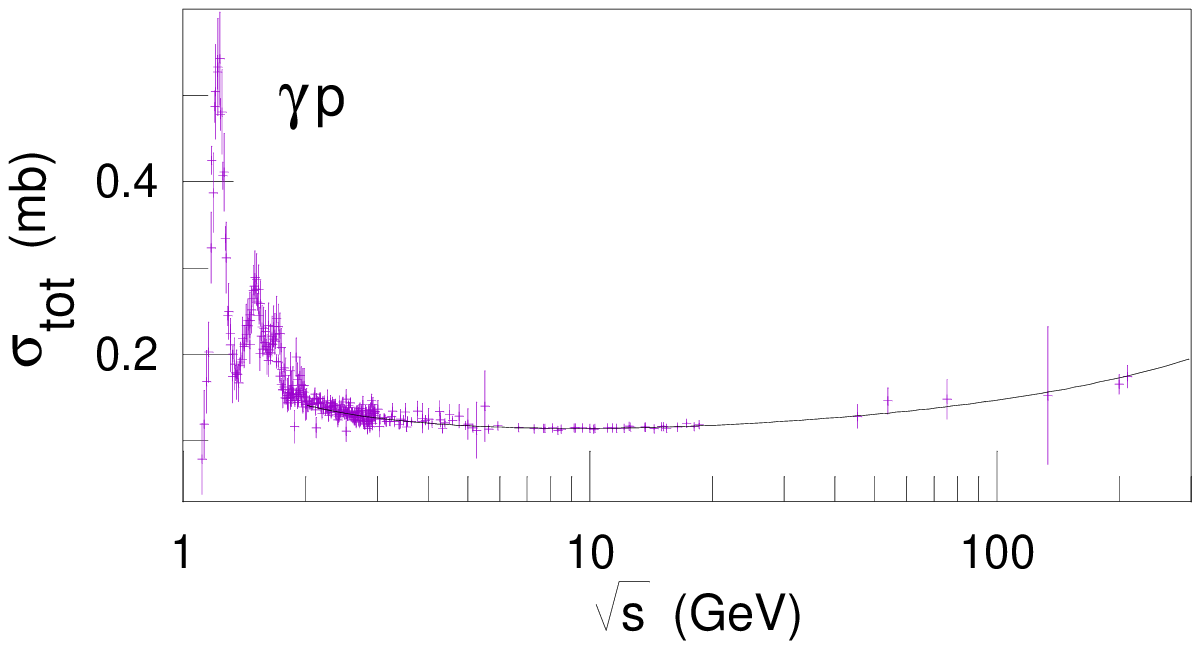}}
\hglue 1cm\resizebox*{0.45\textwidth}{!}{\includegraphics{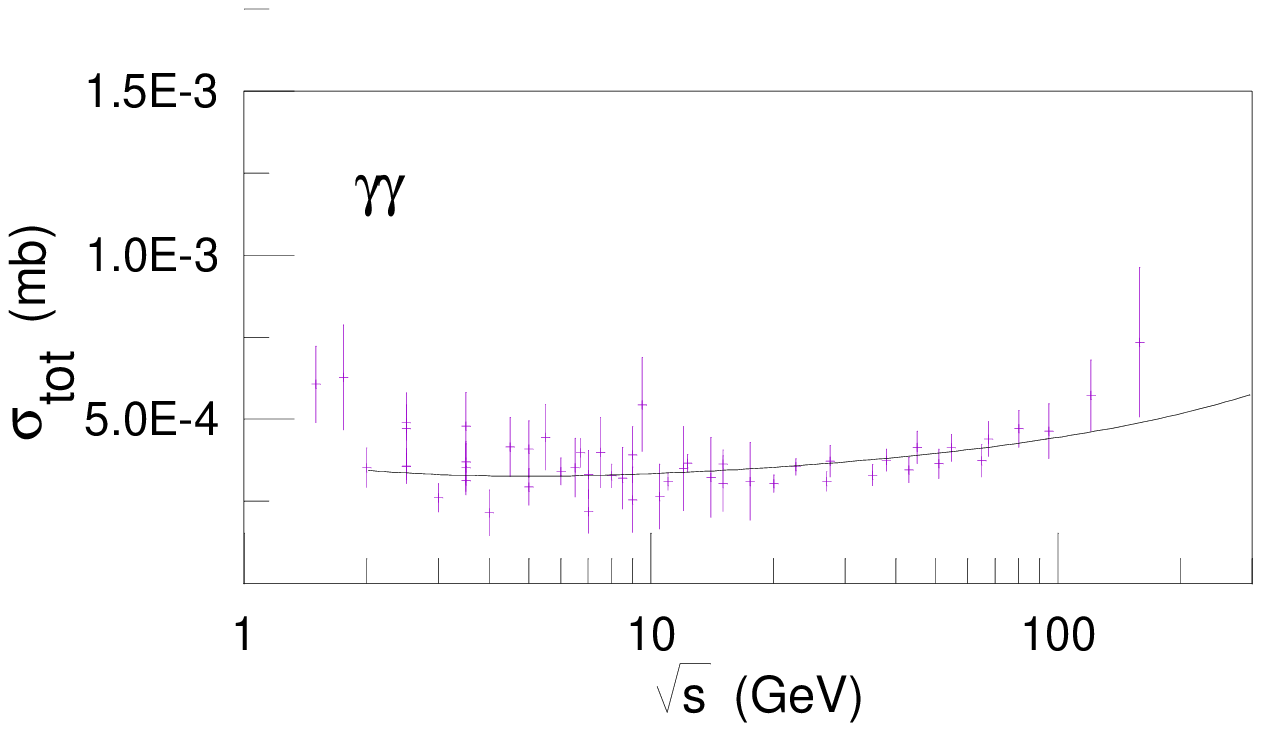}} } \\
Figure 5: Fit to \( \gamma p \) total cross section, and prediction
of \( \gamma \gamma  \) via factorisation. The pole and the unitarised fits 
are identical in the energy range shown.
\vspace{0.3cm}

\subsection{Unitarised fit}

As we have pointed out above, the hard singularity cannot be extended
to energies beyond a few hundred GeV, as one will reach the black-disk
limit in that region, and hence one will have to unitarise the exchange.
The problem of course is that nobody knows how to unitarise Regge
exchanges unambiguously. 

Unitarisation comes from the consecutive exchange of trajectories.
We know that if 1-pomeron exchange is given by the amplitude

\[
\Im mA(s,t)\approx g_{1}\left( \frac{s}{s_{1}}\right) ^{\alpha _{h}}\,
e^{R^{2}t}\]
with \[
R^{2}=B+\alpha '\log s\]
then, if the hadrons remain intact during multiple exchanges, the
\( n \)-pomeron contribution will be proportional to

\[
\Im mA^{(n)}(s,t)=(-1)^{n-1}g_{n}\, s\, \frac{s^{n(\alpha _{h}-1)}}{\left[
R^{2}\right] ^{n-1}}\, e^{\frac{R^{2}}{n}t}\]
To this, one must add the contribution of inelastic channels, or equivalently
that from n-reggeon vertices, which are a priori unknown. Even worse,
the coefficients \( g_{n} \) are also unknown in general. For the
scattering of structureless objects (as in QED or in potential scattering),
one can derive at high energy that \( g_{n}=1/(2^{n-1}nn!) \), which leads
to the eikonal formula. However, both hadrons and reggeons have a
structure, hence it is very likely that this formula is not a good
approximation to the true amplitude. Finally,
in the case of several trajectories, one must take into account mixed
exchanges (e.g. Reggeon-pomeron, etc.).

Hence we present here a possible model that would
lead to a simple-pole picture below 100 GeV, and to a unitarised picture
(for the hard pomeron) at higher energies 
(which is similar to that obtained in the \( U \)-matrix formalism
of \cite{U}). 
As explained above, it is by no means
unique, and many improvements or modifications can be brought in.
Its purpose is not to solve unitarisation,
but only to show that it is possible to accommodate a hard pomeron with
\( t=0 \) data up to the Tevatron\footnote{%
Building of a unitarisation model will necessitate considerable work,
and the adjunction of data at \( t\neq 0 \).
}. 
The simplest choice is to replace (\ref{hardpole})
in Eq. (\ref{hardpompole}) by:\begin{equation}
\label{hardpomunit}
\Im mA^{H}_+(s)=H_{a}sR^{2}\left[ \frac{1}{G}\log \left\{ 1+G\frac{s^{\alpha
_{h}-1}}{R^{2}}\right\} \right] .
\end{equation}
(we shall use again $B=4$ GeV$^{-2}$ and $\alpha'=0.1$ GeV$^{-2}$ in $R^2$). 
 To simplify further, we have assumed that \( G \) would take the
same value \( G_{p} \) for \( p \), \( \pi  \) and \( K \), and
allowed it to be different (and called it \( G_{\gamma } \) )for
\( \gamma p \).

For small values of \( G \), this form reduces to a simple-pole parametrisation
at low energy, and obeys the Froissart bound at high energy. One can
see in Fig. 2 that the simple-pole fit to 100 GeV and the unitarised
fit to 1800 GeV are very close (in fact the \( \log  \) in (\ref{hardpomunit})
and its Taylor expansion to order \( G \) differ by 7\% at \( \sqrt{s}=100 \)
GeV).

Such a form produces the best fit so far to soft data, and we show
the corresponding parameters in Table 3. It clearly can accommodate
the Tevatron data, where the cross section is predicted to be 75.5
mb, and where the hard pomeron contributes about 10\% to the total
cross section. As we pointed out above, this is only a possibility:
we do not know how to unitarise these exchanges, and we assumed that
one could unitarise the hard pomeron independently from the other
exchanges, which is far from clear.

It is worth pointing out that we have fixed the hard and soft pomeron
intercepts to their values measured at lower energies. If we let them
free, then the soft pomeron intercept moves to 1 and the hard pomeron
intercept grows to larger values, but the change in \( \chi ^{2} \)
is not very significant: in fact, the unitarised fit has too many
parameters to be sufficiently constrained by the forward data alone.

\section{Conclusion}

Due to its simplicity and theoretical appeal, the simple pole model
has become quite popular. However, it was shown \cite{COMPETE02}
that it could not accommodate forward scattering data as well as other
fits based on unitary forms. We showed here that the ingredient needed
to restore the simple-pole model as one of the best descriptions \( - \)
besides a careful usage of dispersion relations and the 
lifting of the degeneracy of the \( C=+1 \) and \( C=-1 \)
trajectories \( - \) is precisely the hard pomeron introduced in
DIS\footnote{%
Note that we have also shown that the parametrisation using both soft
and hard pomerons is not the only possible answer: unitary forms
can also provide good fits 
to \( \rho  \) and \( \sigma _{tot} \) \cite{COMPETE02}
(and to elastic slopes \cite{Block} 
and DIS data \cite{CMS}).
}.

Such a hard object cannot be directly observed at high energy, because
it must first be unitarised. However, if one stays below energies
of 100 GeV, the improvement brought in by such a singularity is clearly
visible. We have also shown that it is possible to find unitarised
forms that look like a simple hard pole at low energy, and like a
squared logarithm of \( s \) at high energy. The coupling of the
hard pomeron to protons turns out to be a factor 2 to 3 lower than
that to pions and kaons, whereas that to photons is roughly \( \alpha /\pi  \)
times the coupling to pions.

Hence there are two major questions raised by this possibility of
a hard pomeron in soft data: how does one unitarise the amplitude, especially
in the region of \( \sqrt{s} \) from 100 GeV to the Tevatron, and
why are protons different? Precision data in \( pp \) scattering
in the region from 100 to 600 GeV would have been invaluable in settling
this question. New measurements of \( \rho _{pp} \) would also have
helped decide if the high value of the \( \chi ^{2}/\)n.o.p. for this
observable can be attributed to errors in the data.

One place where one should be able to decide whether the hard pomeron
really exists in soft processes is in \( \gamma \gamma  \) scattering.
If the hard pomeron is present in soft data, then from its contribution
to \( pp \) and to \( \gamma p \), one can predict the \( \gamma \gamma  \)
cross section, both for on-shell and for off-shell photons, and the
presence of a hard pomeron should be manifest in higher-precision
data on the photon-photon and photon-proton total cross sections.

\section*{Acknowledgements}
We thank V.V. Ezhela for comments and corrections, P.V. Landshoff
for discussions and communications, and B. Nicolescu and
K. Kang for discussions about dispersion relations.
E.M. and O.V.S. are visiting fellows of the Fonds National de la Recherche
Scientifique (F.N.R.S.). A.L. thanks the F.N.R.S. and the Université
de Liège, where this work was done, for their support and hospitality.

\hfill{}
\end{document}